\documentstyle[12pt,agums]{article}

\lefthead{Summers and Ma}
\righthead{
acceleration of Electrons by Fast-Mode waves
}
\received{November~5,~1999}
\revised {Febrary~4,~2000}
\accepted{February~23,~2000}
\paperid{1999JA000408}
\cpright{AGU}{1999}
\ccc{0148-0227/00/1999JA000408\$09.00}
\authoraddr{Danny Summers and Chun-yu Ma,\\
Department of Mathematics and Statistics,
Memorial University of Newfoundland,\\
St John's, Newfoundland, A1C 5S7, Canada
(e-mail: dsummers@math.mun.ca, cyma@math.mun.ca)}

\slugcomment{
}

\begin{document}

\title{
RAPID ACCELERATION OF ELECTRONS IN THE MAGNETOSPHERE BY FAST-MODE MHD
WAVES
}

\author{Danny Summers and Chun-yu Ma \altaffilmark{1} 
        }
\bigskip

\affil{Department of Mathematics and Statistics, Memorial University of
Newfoundland,\\ St John's, Canada}

\altaffiltext{1}{
On leave from Purple Mountain Observatory, Chinese Academy of Sciences,
Nanjing, People's Republic of China.}  

\begin{abstract}
During major magnetic storms enhanced fluxes of relativistic
electrons in the inner magnetosphere have been observed to correlate with
ULF waves. The enhancements can take place over a period of several hours.
In order to account for such a rapid generation of relativistic electrons,
we examine the mechanism of transit-time acceleration of electrons by
low-frequency fast-mode MHD waves, here the assumed form of ULF waves.
Transit-time damping refers to the resonant interaction of electrons with
the compressive magnetic field component of the fast-mode waves via the
zero cyclotron harmonic. In terms of quasi-linear theory, a kinetic
equation for the electron distribution function is formulated
incorporating a momentum diffusion coefficient representing transit-time
resonant interaction between electrons and a continuous broad band
spectrum of oblique fast-mode waves. Pitch angle scattering is assumed to
be sufficiently rapid to maintain an isotropic electron distribution
function. It is further assumed that there is a substorm-produced
population of electrons with energies of the order of 100 keV.
Calculations
of the acceleration timescales in the model show that fast-mode waves in
the Pc4 to Pc5 frequency range, with typically observed wave amplitudes
($\Delta B=10$--$20$ nT), can accelerate the seed electrons to energies
of order MeV in a period of a few hours. It is therefore concluded that
the mechanism examined in this paper, namely, transit-time acceleration of
electrons by fast-mode MHD waves, may account for the rapid enhancements
in relativistic electron fluxes in the inner magnetosphere that are
associated with major storms.
\end{abstract}

\begin{article}

\section{Introduction}

There is much current interest in the rapid enhancements of relativistic
($>$MeV) electrons in the Earth's inner magnetosphere ($3 \leq L \leq 6$)
taking place over tens of minutes or a few hours 
during major magnetic storms [e.g.,
\markcite{{\it Baker et al.}, 1998a; {\it
Rostoker et al.}, 1998; {\it Liu et al.}, 1999; {\it Hudson et
al.}, 1999,2000}].
Part of this interest is due to the fact that relativistic
electrons appearing near geostationary orbit ($L=6.6$) constitute a
potential hazard to operational spacecraft [e.g.,
\markcite{{\it Baker et al.,} 1997}].
These relativistic electrons are sometimes colloquially referred to as
``killer electrons."
Rapid energetic electron
enhancements have been observed to correlate closely
with ULF waves in the Pc4 (7--22 mHz) or Pc5 (2--7 mHz) frequency ranges.
It is therefore reasonable to examine the possible role that ULF waves may
have in generating the relativistic electron flux enhancements.
\markcite{{\it Liu et al.} [1999]} have formulated an acceleration
mechanism comprising magnetic
pumping by ULF waves, while
\markcite{{\it Hudson et al.} [1999,2000]} have proposed a
drift-resonant acceleration
mechanism involving enhanced ULF waves, modeled by a three dimensional 
global MHD simulation of the
January 10-11, 1997, conoral-mass-ejection-driven magnetic cloud event.
Notwithstanding these
studies, the acceleration mechanism of relativistic electrons in the inner
magnetosphere is not yet fully understood. It is the purpose of the present
paper to examine the role of ULF waves in accelerating electrons in the
magnetosphere from a new standpoint. Here we examine the ``transit-time
acceleration" of electrons by low-frequency, oblique, fast-mode
(magnetosonic) MHD waves. Transit-time acceleration in association with
``transit-time damping" has been studied, for instance, by
\markcite{{\it Stix} [1962], {\it Fisk} [1976]}, and
\markcite{{\it Achterberg} [1981]}.
The basic physical mechanism of transit-time damping, which is a resonant
form of Fermi acceleration and can be regarded as the magnetic analogue of
Landau damping, is discussed in detail by
\markcite{{\it Miller} [1997]};
the name transit-time damping arises because the gyroresonance condition
defining the process can be expressed as
$\lambda_\parallel/v_\parallel\approx T$, where $v_\parallel$ is the
parallel component of particle velocity,
$\lambda_\parallel$ is the parallel wavelength, and $T$ is the wave period.
Thus the wave-particle interaction is strongest when the particle
transit-time across the wave compression is approximately equal to the period.
It is the compressive magnetic field component of the fast-mode wave that
allows for the effect of transit-time damping
\cite{fisk76,achter81,miller96}.
While transit-time damping has been utilized as a mechanism for accelerating
energetic particles in the interplanetary medium
\cite{fisk76},
for accelerating electrons in solar flares
\cite{miller96,miller97},
and for accelerating cosmic ray particles
\cite{sch98},
it has not been examined as a possible acceleration mechanism of electrons in
the magnetosphere. It will be shown in this paper, in fact,
that transit-time damping of fast-mode MHD waves 
(here the assumed form of ULF waves)
is a viable mechanism for
generating the aforementioned rapid enhancements of relativistic electrons
in the inner magnetosphere. It will be assumed that electrons of energies
$\sim$100 keV that are injected near geosynchronous orbit as a result
of substorm activity [e.g.,
\markcite{{\it Baker et al.}, 1989, 1998b}]
form the source population for the relativistic ($>$MeV) electrons that
are subsequently observed.

The structure of ULF waves in the Earth's magnetosphere is
complex. Broadly, MHD waves in the dipole magnetosphere are
characterized by ``toroidal" and ``poloidal" modes, and, in general, these
modes
are coupled. Toroidal modes relate to transverse
ULF waves propagating along field lines, while poloidal modes relate
to global compressional waves associated with radial oscillations of
the field lines. Field line resonance (FLR) theory describes the toroidal
pulsations as transverse Alfv\'en waves standing on dipole flux tubes
with fixed ends in the ionosphere. Extensive theory has accumulated
pertaining to field line resonances, global compressional modes, and
associated wave excitation mechanisms [e.g., see
\markcite{{\it Kivelson and Southwood}, 1986}, 
\markcite{{\it Krauss-Varban and Patel}, 1988},
\markcite{{\it Lee and Lysak}, 1989}, and references therein].
\markcite{{\it Anderson et al.} [1990]} give a historical review of 
observations of ULF waves in the magnetosphere and also present
results of a statistical study of Pc3--5 pulsations during the period
from August 24, 1984, to December 7, 1985. It should be emphasized here
that unlike ULF waves observed during relatively quiet magnetic
conditions, storm-associated ULF waves characteristically have large
compressional components. It is these components that engender the
transit-time acceleration mechanism presented in this paper. 

With respect to the typical motion of energetic electrons in the inner
magnetosphere, these electrons gyrate around their magnetic field line
and bounce back and forth along their field line between
mirror points, while executing an eastward drift about the
Earth. Consequently, since in the interests of rendering the theory
presented herein tractable we shall assume a constant background
magnetic field, our formulation applies to compressional MHD waves
interacting with electrons that mirror relatively close to the equator.

We are concerned with the fast-mode MHD branch in plasma wave theory
[e.g.,
\markcite{{\it Swanson}, 1989}]
and, in particular, with fast-mode waves on the $\omega \ll \Omega_i$
section of the branch,
where $\Omega_i$ is the proton gyrofrequency.
Such waves have the dispersion relation,
\begin{equation}
   \omega=kv_A,                         
\end{equation}
where $\omega, \, k$, and $v_A$ are the wave frequency, the wave number,
and the Alfv\'en
speed, respectively. In general, the condition for gyroresonant interaction
between electrons and a wave of frequency $\omega$ is
\begin{equation}
          \omega-k_\parallel v_\parallel=n|\Omega_e| /\gamma,
\end{equation}
where $v_\parallel$ is the electron parallel velocity component ,
$k_\parallel=k\cos\theta$
is the parallel wave number,  $\theta$ is the wave propagation angle,
$|\Omega_e|$
is the electron gyrofrequency, $\gamma$ is the Lorentz factor, and
$n \, (=0, \, \pm 1, \, \pm 2, \, \cdots)$ denotes the cyclotron harmonic.
The
compressive component of the wave magnetic field can interact with electrons
through the $n=0$ resonance [e.g.,
\markcite{{\it Miller et al.}, 1996}], in which case (2)
reduces to
\begin{equation}
   \omega=k_\parallel v_\parallel.
\label{3}
\end{equation}
Equation (\ref{3}) is the gyroresonance condition that defines
transit-time
damping 
\cite{stix62,fisk76,achter81}.
From (1) and (3) it follows that
\begin{equation}
  v_\parallel=v_A/\cos\theta,
  \label{4}
\end{equation}
from which follows the important necessary threshold condition for resonance,
\begin{equation}
  v>v_A,
  \label{5}
\end{equation}
where $v$ is the particle speed.                    
Condition (\ref{5}) states that for electrons with any pitch-angle
interacting with fast-mode MHD waves propagating at any angle $\theta$,
resonance is only possible
for electrons with speeds exceeding the Alfv\'en speed. The equivalent
minimum-energy condition can be conveniently written,
\begin{eqnarray}
 & E > E_{\min} \, ,& \nonumber\\
 & E_{\min} = (1-\beta_A^2)^{-1/2}-1 \approx \beta_A^2/2 \, ,&
\label{6}
\end{eqnarray}
where $E$ is the electron kinetic energy in units of rest-mass energy and
$\beta_A$ is the Alfv\'en speed in units of the speed of  light. Values
for
the parameter $\beta_A$ and the minimum energy $E_{min}$
that are representative of the inner
magnetosphere are given in 
\callout{Table 1\ref{table1}}. 
In Table 1\ref{table1},
we set $N_0=10$ cm$^{-3}$ as the particle number density in the inner
magnetosphere outside the plasmasphere, and we use the equatorial (dipole)
magnetic field value $B_0=3.12\times 10^{-5}/L^3$ T. With regard to the
background electron population, in order for the fast-mode waves to 
accelerate a small fraction of the electrons in the tail
of the distribution rather than to produce a bulk heating of the
population, it is
required that $v_A>v_{th}$, where $v_{th}$ is a characteristic thermal
speed. Taking
the background electron temperature in the magnetosphere to be $T_e
\widetilde{<}1$ eV,
we find from Table 1\ref{table1} that the required condition $v_A>v_{th}$
is satisfied. In addition, since we are assuming a substorm-produced source
of electrons with energies $\sim$100 keV, Table 1\ref{table1} shows that
the
minimum-energy condition (\ref{6}) is well satisfied; that is, the
condition $v \gg v_A$ holds.

The analysis of transit-time damping of fast-mode waves that is carried
out in this
paper and presented in the following section is based on quasi-linear theory;
this is an approximation that requires justification. In numerical simulations,
\markcite{{\it Miller} [1997]} has found that quasi-linear theory
provides an accurate
description of transit-time acceleration even when the energy density
of the fast-mode wave turbulence is almost equal to the ambient magnetic field
energy density $((\Delta B/B_0)^2 \widetilde{<} 1)$. 
Thus, although the analysis presented here is based formally
on ``small-amplitude" turbulence $((\Delta B/B_0)^2 \ll 1)$, 
the results are applicable to the
large-amplitude ULF waves typically observed during magnetic storms.

\section{Electron Momentum Diffusion Equation}

Consider energetic charged particles in a uniform magnetic field with
superimposed small-amplitude plasma turbulence. By using the quasi-linear
approximation 
\cite{ken66,lerche68},
the pitch angle averaged particle distribution function $F(p,t)$ can be
shown
to satisfy the kinetic (Fokker-Planck) equation
\begin{equation}
  \frac{\partial F}{\partial t} =
  \frac{1}{p^2} \frac{\partial}{\partial p}
    \left( p^2D(p) \frac{\partial F}{\partial p} \right),
\label{7}
\end{equation}
where
\begin{equation}
  D(p)=\frac{1}{2} \int_{-1}^1 D_{pp} d\mu.
\label{8}
\end{equation}
In (\ref{7}) and (\ref{8}), $p$ is the relativistic momentum of the
particle in units of rest-mass momentum
given by $p=\gamma v/c$, where $v$ is the particle speed and $\gamma=
(1-v^2/c^2)^{-1/2}=(1+p^2)^{1/2}$ is the Lorentz factor, with $c$ being 
the speed
of light; $t$ is time; $\mu$ is the cosine of the pitch angle; and
$D_{pp}$
is the momentum diffusion coefficient, which depends on the properties of the
wave turbulence. In the derivation of (\ref{7}), it has been assumed
that the rate of pitch angle scattering is large enough to isotropize the
distribution function, and the pitch angle  has been eliminated from the equation
by averaging with respect to $\mu$. The distribution function $F$ is
normalized so that  $4\pi p^2F(p,t) dp$ is the number of the particles per unit
volume in the momentum interval $dp$. It has also been assumed in deriving
(\ref{7}) that there are no energy losses, that no particles escape from the
system, and that there are no additional particle sources or sinks.

Associated with the (averaged) momentum diffusion coefficient $D(p)$ in
(\ref{7}) and (\ref{8}) is the acceleration timescale,
\begin{equation}
  T_A=p^2/D(p).
  \label{9}
\end{equation}
In this paper we consider two forms for the transit-time damping
diffusion coefficient, given by
\markcite{{\it Miller et al.} [1996]} and
\markcite{{\it Schlickeiser and Miller} [1998]}, respectively.
Assuming a continuous spectrum of oblique, low-frequency
($\omega \ll \Omega_i$), fast-mode waves and assuming isotropic turbulence
and integrating over wave propagation angle,
\markcite{{\it Miller et al.} [1996]} obtain a diffusion coefficient
$D_{pp}$ for transit-time
damping of fast-mode waves by electrons that can be expressed in the form
\begin{equation}
  D_{pp}=\frac{\pi}{16} \Omega_i R
         \left( \frac{c\langle k \rangle}{\Omega_i} \right)
         \gamma^2\beta\beta_A^2
         \left( 1-\frac{\beta_A^2}{\beta^2\mu^2} \right)
         \frac{(1-\mu^2)^2}{|\mu|} \; ,
 \label{10}
\end{equation}
where $\Omega_i=eB_0/(m_ic)$ is the proton gyrofrequency, with $B_0$
being the
ambient magnetic field strength, $m_i$ being the proton rest mass, and $e$
being the electronic charge; $R=(\Delta B/B_0)^2$ is the ratio of the
turbulent
wave energy to magnetic field energy, with $\Delta B$ being the average
fast-mode
wave amplitude; $c\langle k \rangle /\Omega_i$ is the mean dimensionless
wave number of the wave spectrum; $\beta=v/c$; and $\beta_A=v_A/c$ where
$v_A=B_0/(4\pi N_0m_i)^{1/2}$ is the Alfv\'en speed, with $N_0$ being the
particle
number density. Substituting (\ref{10}) into (\ref{8}) and setting
\begin{equation}
  x=\beta_A/\beta
\end{equation}
yields the result
\begin{equation}
  D(p)=\frac{\pi}{16} \Omega_i R
         \left( \frac{c\langle k \rangle}{\Omega_i} \right)
         \gamma^2\beta\beta_A^2 g(x) \; ,
 \label{12}
\end{equation}
where
\begin{equation}
g(x)=(1+2x^2)\log_e\left(\frac{1}{x}\right)+x^2+\frac{x^4}{4}-\frac{5}{4} 
\; ,
\end{equation}
for $x<1$.
The function $g(x)$ can be regarded as an efficiency factor 
\cite{miller96}, which relates to the velocity-dependent fraction of
electrons that can resonate with fast-mode waves having the assumed spectrum;
$g(x)=0$ for $x\geq 1$, and in the limit as $\beta \rightarrow 1$,
$g(x)\rightarrow \log_e(1/\beta_A)-5/4$, approximately, since $\beta_A \ll
1$.
Therefore, for values of $\beta_A$ appropriate to the Earth's magnetosphere
(see Table 1\ref{table1}) for $3\leq L\leq 6.6$, with $N_0=10$ 
cm$^{-3}$, the
function $g(x)$ approaches values in the range from 2.4 to 4.7 for highly
relativistic electrons.

Setting $\langle k \rangle =\langle \omega \rangle /v_A$ in (\ref{12}), where
$\langle \omega \rangle$ is the mean angular frequency (rad/s), from
(\ref{9}) and (\ref{12}) we find that the acceleration timescale $T_A$ can be
written as
\begin{equation}
  T_A=\frac{8}{\pi^2} \frac{1}{\langle f_w \rangle} \frac{1}{R}
       \frac{1}{xg(x)} \; ,
 \label{14}
\end{equation}
where $\langle f_w \rangle=\langle \omega \rangle/2\pi$ is the mean wave
frequency (in millihertz). Later in this section and in the numerical
results presented
below, we shall find it convenient to use the previously
introduced dimensionless kinetic energy
$E=E_k/(m_ec^2)=\gamma-1$, where $E_k$ is the electron kinetic energy and
$m_e$ is the electron rest mass; we shall require the relation,
\begin{equation}
  \beta=\left[ E(E+2) \right]^{1/2}/(E+1).
  \label{15}
\end{equation}

\markcite{{\it Schlickeiser and Miller} [1998]} assume that the
fast-mode wave turbulence is isotropic and
Kolmogorov-like, with a power law spectral energy density distribution in
wave number $k$. Specifically, the spectral energy density $W$ is assumed
to take the form
\begin{equation}
   W(k)\propto k^{-q}, \; k>k_{min},
   \label{16}
\end{equation}
where $q(>1)$ is the spectral index and $k_{min}$ is some minimum
wave number.
Corresponding to the Kolmogorov-like spectrum (\ref{16}), the momentum diffusion
coefficient $D_{pp}$, as given by
\markcite{{\it Schlickeiser and Miller} [1998]}, can be written
\begin{equation}
  D_{pp}=\frac{\pi}{4}(q-1)\Omega_i R
           \left( \frac{ck_{min}}{\Omega_i} \right)^{q-1}
           \left( \frac{m_e}{m_i} \right)^{q-2} \gamma (\gamma\beta)^{q-1}
           \beta_A^2 h(\mu,x),
  \label{17}
\end{equation}
where
\begin{equation}
  h(\mu,x)=H(|\mu| -x)
           \frac{1-\mu^2}{|\mu|}
           \left[ 1+\frac{x^2}{\mu^2} \right]
           \left[ (1-\mu^2)\left(1-\frac{x^2}{\mu^2}\right) \right]^{q/2}
           \int_\lambda^\infty \frac{J_1^2(s)}{s^{1+q}} ds,
  \label{18}
\end{equation}
with
\begin{equation}
  \lambda=\left( \frac{ck_{min}}{\Omega_i} \right)
          \left( \frac{m_e}{m_i} \right)
          \gamma\beta (1-\mu^2)^{1/2}
          \left( 1-\frac{x^2}{\mu^2} \right)^{1/2}.
  \label{19}
\end{equation}
In (\ref{17})--(\ref{19}), $ck_{min}/\Omega_i$ is the minimum dimensionless
wave number of the wave spectrum, $H$ is the Heaviside unit function, and
$J_1$
is the Bessel function of the first kind of order unity. Substitution of
(\ref{17}) into (\ref{8}) yields
\begin{equation}
  D(p)=\frac{\pi}{4}(q-1)\Omega_i R
          \left( \frac{ck_{min}}{\Omega_i} \right)^{q-1}
          \left( \frac{m_e}{m_i} \right)^{q-2}
          \gamma (\gamma\beta)^{q-1} \beta_A^2
          I(x,\beta_A,k_{min}),
  \label{20}
\end{equation}
where
\begin{equation}
  I(x,\beta_A,k_{min})=
  \cases{
          c_1(q)\log_e\left(\frac{1}{x}\right), ~~~~~ 1<q\leq 2 \cr
          c_2(q)(\gamma\beta)^{2-q}
             \left( \frac{ck_{min}}{\Omega_i} \right)^{2-q}
             \left( \frac{m_e}{m_i} \right)^{2-q}
             \log_e\left(\frac{1}{x}\right), ~~~~~~~~~~ q>2
        }
  \label{21}
\end{equation}
with
\begin{eqnarray}
  c_1(q) & = & 2^{1-q} \frac{q}{4-q^2}
                \frac{\Gamma(q)\Gamma(2-q/2)}{\Gamma^3(1+q/2)}, ~~~~~ 1<q<2 
\nonumber\\
  c_1(2) & = & 3/4,  \\
  c_2(q) & = & \frac{2q^2-3q+4}{4q(2q-3)}, ~~~~~~~~~~~~~~~ q>2 \nonumber
\end{eqnarray}
where $\Gamma$ is the gamma function.

In (\ref{20}) we set the minimum dimensionless wave number
$k_{min}=2\pi f_{min}/v_A$, where $f_{min}$ is the minimum wave frequency
(in millihertz). From (\ref{9}) and (\ref{20}) the acceleration timescale
for
transit-time damping associated with the wave spectrum (\ref{16}) is found
to be
\begin{equation}
  T_A=\frac{8}{q-1} \frac{1}{(2\pi)^q} \frac{1}{\Omega_i} \frac{1}{R}
          \left( \frac{\Omega_i}{f_{min}} \right)^{q-1}
          \left( \frac{m_i}{m_e} \right)^{q-2}
          \frac{1}{\gamma^{q-2}} \frac{1}{x^{3-q}I},
  \label{23}
\end{equation}
where $I$ is given by (\ref{21}) with $ck_{min}/\Omega_i$ replaced by
$(f_{min}/\Omega_i)(2\pi/\beta_A)$.

It should be noted that while the transit-time damping diffusion
coefficients (12) and (20) may appear different, they are, in fact,
approximately equivalent. Since the coefficient (12) employs an
average wave frequency, while coefficient (20) employs a minimum
frequency and a Kolmogorov spectral index, it is convenient to utilize
both coefficients in order to retain some flexibility in constructing
the acceleration timescale profiles and comparing the results with
observations.

Finally, it is useful to relate the mean energy change of a particle
$\langle \dot{E} \rangle$ to the acceleration timescale $T_A$.
Associated with the momentum diffusion process given by (\ref{7}) --
(\ref{8}), the mean energy change 
\cite{tsy77,achter81}
is given by
\begin{eqnarray}
  \langle \dot{E} \rangle &=& \frac{1}{p^2}
        \frac{\partial}{\partial p}
          \left( \beta p^2D(p) \right) \nonumber \\
                         & \approx &
       \frac{\sigma\beta}{p}D(p),
 \label{24}
\end{eqnarray}
and, hence, by (9) we derive the result,
\begin{equation}
 \langle \dot{E} \rangle \approx \frac{\sigma\beta p}{T_A} =
        \frac{\sigma E(E+2)}{(E+1)T_A},
 \label{25}
\end{equation}
where $\sigma$ is a factor such that $\sigma=4$ corresponding to the
diffusion coefficient (\ref{12}), and $\sigma=2+q$ corresponding to
(\ref{20}). The approximation in the second line of (24) follows from
the fact that, for the electron energies considered in this paper, the
functions $g$ and $I$ vary only slightly with $x$.

\section{Numerical results}

The acceleration timescale $T_A$ depends on a number of parameters. Both
results (\ref{14}) and (\ref{23}) depend on the average wave amplitude
$\Delta B$, the electron kinetic energy $E$, the background plasma number
density $N_0$, and the location $L$. In addition, (\ref{14}) depends on
the mean wave frequency $\langle f_w \rangle$, while (\ref{23}) depends on
the the minimum wave frequency $f_{min}$ and the turbulence  spectral index
$q$. With regard to
typical wave amplitudes of Pc-5 pulsations during major magnetic storms,
\markcite{{\it Barfield and McPherron} [1978]} and
\markcite{{\it Engebretson and Cahill} [1981]} report $\Delta
B\approx 10$ nT, while
\markcite{{\it Higuchi et al.} [1986]} report typical values $\Delta B
\approx$ 70 -- 90 nT corresponding to the maximum power spectral
densities in the frequency range 5 -- 12 mHz.
\markcite{{\it Baker et al.} [1998{\it a}]} report ULF waves in the
frequency range 2 --20 mHz having
amplitudes $\Delta B\approx 50$ nT, rising to $\Delta B\approx 200$ nT
at times.

We assume a substorm-produced seed electron population with energies
in the range 100 -- 300 keV, which corresponds to the dimensionless 
kinetic energy $E$ in the
approximate range $0.2 < E < 0.6$. From result (\ref{25}), it follows that
electrons with energies in such a range accelerate to energies in the
range from 1 MeV to 2 MeV, approximately, over the timescale $T_A(E)$
where $0.2 < E < 0.6$. In Figure ~\ref{fig1}, 
$N_0=10$ cm$^{-3}$, and for the specified mean wave frequency $\langle f_w
\rangle = 10$ mHz, curves are plotted showing $T_A$, given by (\ref{14}),
as a function of energy $E$ (eV), at the locations $L=3,\,4,\,5,\,6.6$,
for each of the wave amplitudes $\Delta B=10,\,20,\,50$ nT; for reference,
a (dashed) line is shown corresponding to a time of one day. In general,
for a
fixed energy $E$, the timescale $T_A$ is seen to increase as the value of
$L$ decreases, and, as expected, $T_A$ decreases as the wave amplitude
$\Delta B$ increases. Figure ~\ref{fig1} indicates, in particular, that at
$L=6.6$, for the parameter values $N_0=10$ cm$^{-3}$ and $\langle f_W
\rangle = 10$ mHz, the timescales for accelerating seed electrons of
energies  $\sim 100$ keV to energies  $\sim 1$ MeV are approximately 6
days, 1.5 days, and 5.8 hours corresponding to the respective wave
amplitudes $\Delta B=10,\,20,\,50$ nT. The aforementioned respective times
assume the approximate values 2 days, 12 hours, and 2 hours if the value
of $N_0=1$ cm$^{-3}$ is specified for the background plasma number
density (N.B. It could be argued that $N_0=10$ cm$^{-3}$ is too high a
generic value for the background plasma number density, and that $N_0=1$
cm$^{-3}$ is a more representative value). 

In Figure \ref{fig2}, we show the variation of $T_A$, as given by
(\ref{14}), as a function of the wave amplitude $\Delta B$ (nT), for a
fixed value of particle
energy, $E=1/4$ (or $\beta=0.6$). The upper, middle,
and lower panels of Figure ~\ref{fig2} correspond respectively to the
mean wave frequencies $\langle f_w \rangle=2,\,10$, and $22$ mHz. The
decrease in acceleration timescale with increase in mean frequency
$\langle f_w \rangle$, 
as indicated by formula (\ref{14}), is clearly
shown in Figure ~\ref{fig2}. 
Figure ~\ref{fig2} can be used as an illustration of the wave
amplitudes $\Delta B$, at a given location $L$, and for a given
mean wave frequency $\langle f_w \rangle$, that correspond to a particular
timescale $T_A$ for the generation of electrons of energies $\gtrsim 1$
MeV from seed electrons of energies $\gtrsim 100$ keV. 
In Table 1~\ref{table1}, 
corresponding to (\ref{14}), the required wave amplitudes
$\Delta B$ are given that correspond to a timescale $T_0=10$ hours for
this generation process, corresponding to the mean wave frequencies 
$\langle f_w \rangle=2,\,10,$ and $22$ mHz at the specified locations, and
with $N_0=10$ cm$^{-3}$. In particular, we note that at $L=6.6$,
corresponding to the respective mean wave frequencies 
$\langle f_w \rangle=10,\,22$ mHz, the required wave amplitudes are 
$\Delta B=39,\,26$ nT; corresponding to $N_0=1$ cm$^{-3}$, these
respective wave amplitudes are $\Delta B= 22, \, 15$ nT.

In Figure ~\ref{fig3}, for $N_0=10$ cm$^{-3}$, the acceleration timescale
$T_A$ (sec) given by (\ref{23}) is plotted as a function of particle
energy $E$, for the minimum wave frequencies $f_{min}=2,\, 10$ mHz, 
for the spectral indices $q=3/2, \, 5/3, \, 5/2, \, 4$, 
at each of the locations $L=3, \, 6.6$, 
and for the mean wave amplitude $\Delta B=20$ nT. As
can be observed from Figure ~\ref{fig3}, the timescale $T_A$ decreases as
both $f_{min}$ and $L$ increase; lower values of $T_A$ are also generally
favoured by lower $q$-values. The curves in Figure ~\ref{fig3} show that,
corresponding to $\Delta B=20$ nT, the timescale $T_A$ at $L=6.6$ is of
the order of a few hours, for values of $q$ in the range $3/2 < q < 5/3$.
In Table 2 ~\ref{table2}, values of the
wave amplitudes $\Delta B$ (nT) are given that correspond to the value
of the acceleration timescale $T_A$ given by
(\ref{23}) equal to  10 hours, for the specified values of $f_{min}$, $q$,  
and $L$, with $E=1/4$ (or $\beta=0.6$), and $N_0=10$ cm$^{-3}$. Thus,
Table 2 ~\ref{table2}, which corresponds to (\ref{23}), effectively gives
the required wave amplitudes $\Delta B$ to generate relativistic 
($\gtrsim 1$ MeV)
electrons from seed ($\gtrsim 100$ keV) electrons in a timescale of 10
hours, for the specified values of the remaining parameters. For instance,
at $L=6.6$, for a minimum wave frequency $f_{min}=10$ mHz, and with $q$ in
the range $3/2 <q < 5/3$, the required wave amplitudes are in the range 
$5.4$ nT $< \Delta B < 7.1$ nT.

\section{Discussion}

The present paper is a new examination of ULF waves as  a possible rapid
acceleration mechanism of electrons in the inner magnetosphere during
storms. Specifically, we take the assumed form of  
ULF waves to be fast-mode (magnetosonic)
MHD waves, and analyze the mechanism of transit-time acceleration of
electrons under magnetic storm conditions. We assume that the seed
electrons in the process have energies in the range 100 -- 300 keV, and
are produced by substorm activity. In accordance with quasi-linear theory
and a test particle approach, a simple model kinetic equation (\ref{7})
is formulated in which momentum diffusion is due to the gyroresonant
transit-time interaction between electrons and fast-mode MHD turbulence. A
continuous broad-band spectrum of oblique fast-mode waves is assumed,
and it is further supposed that pitch-angle scattering is sufficiently
rapid to maintain an isotropic particle distribution function. The model
calculations applied to the inner magnetosphere show that the mechanism
under
consideration, namely transit-time damping of fast-mode MHD waves, can
accelerate source electrons with energies 100 -- 300 keV to relativistic
electrons with energies exceeding 1 MeV, in a timescale of a few hours if
the wave amplitudes are of the order of $\Delta B=10$ -- 20 nT. Since
observed amplitudes of ULF waves during storm-time are in this range, it
is concluded
that transit-time damping of fast-mode MHD waves, as the agent of ULF wave
activity, could play an important role in generating the observed
increases of relativistic electrons during major storms. 

We note that the models formulated by 
\markcite{{\it Liu et al.} [1999]} and 
\markcite{{\it Hudson et al.} [1999{\it a, b}]}
also show that ULF waves could be instrumental in energizing relativistic
electrons under storm conditions, though their approaches are quite
different from that adopted here. 
\markcite{{\it Liu et al.} [1999]} formulate an acceleration mechanism
comprising magnetic pumping with global ULF waves as the energy source
and pitch-angle scattering as the catalyst, while 
\markcite{{\it Hudson et al.} [1999{\it a, b}]} propose a mechanism,
further investigated by 
\markcite{{\it Elkington et al.} [1999]}, in which electrons are
adiabatically accelerated through a drift-resonance via interaction
with toroidal-mode ULF waves.

We caution that the calculations
in the present paper are based on an approximate, timescale analysis. A
more complete investigation of electron acceleration by transit-time
damping of fast-mode waves entails the full solution of a kinetic equation of
the form (\ref{7}), appropriately modified by the inclusion of terms
representing particle and energy losses under storm conditions. 

Aside from the aforementioned ULF wave mechanisms, other
energization mechanisms have been previously proposed to account 
for relativistic electron enhancements during storms, e.g., see
\markcite{{\it Li et al.} [1997]} and
\markcite{{\it Summers and Ma} [1999]} for brief summaries. Moreover,
various types of storm-related energetic electron events have been
observed [e.g., 
\markcite{{\it Baker et al.}, 1997; 1998{\it b}};
\markcite{{\it Reeves}, 1998}; 
\markcite{{\it Reeves et al.}, 1998]}.
The rapid acceleration mechanism presented in this paper appears well suited
to major storms that produce coherent global oscillations in the
magnetosphere in the Pc-4 to Pc-5 frequency range.  In contrast,
the gradual acceleration process occurring over a few days
involving gyroresonant electron-whistler-mode chorus interaction [
\markcite{{\it Summers et al.}, 1998, 1999};
\markcite{{\it Summers and Ma}, 1999}; see also
\markcite{{\it Ma and Summers}, 1998]} 
is expected to apply to moderate storms having long-lasting
recovery phases.

\acknowledgments
This work is supported by the Natural Sciences and Engineering Research
Council of Canada under Grant A-0621. Additional support is acknowledged
from the Dean of Science, Memorial University of Newfoundland.

{}
 
\end{article}

\clearpage

\begin{figure}
\caption{
Acceleration timescale $T_A$ (sec) as given by (\ref{14}), as a function
of the electron kinetic energy $E$ (eV), at the indicated locations $L$,
for the average wave amplitudes $\Delta B=10,\,20,\,50$ nT. The mean wave
frequency $\langle f_w \rangle=10$ mHz.} 
\label{fig1}
\end{figure}

\begin{figure}
\caption{
Acceleration timescale $T_A$ (sec) as given by (\ref{14}), as a function
of the average wave amplitude $\Delta B$ (nT), at the indicated locations
$L$, for the mean wave frequencies $\langle f_w \rangle=2,\,10,\,22$ mHz.
The parameter $\beta=0.6$.} 
\label{fig2}
\end{figure}

\begin{figure}
\caption{
Acceleration timescale $T_A$ (sec) as given by (\ref{23}), as a function
of the electron kinetic energy $E$ (eV), at the indicated locations $L$,
for the given values of the spectral index $q$,
and for the minimum wave frequencies $f_{min}=2,\,10$ mHz.
The average wave amplitude $\Delta B=20$ nT. }
\label{fig3}
\end{figure}

\clearpage

\begin{table}
\caption{
The required average wave amplitudes $\Delta B$ (nT), as calculated from
(\ref{14}), that correspond to an acceleration timescale $T_A$ of about
10 hours, at the given locations $L$, and for the mean wave frequencies
$\langle f_w \rangle=2,\,10,\,22$ mHz. The parameter $\beta=0.6$. Also
given are the ambient magnetic field strength $B_0$ ($10^{-7}$ T), the
dimensionless Alfv\'en speed $\beta_A=v_A/c$, and the minimum energy
$E_{min}$ (eV). The latter value is calculated from $E_{min}$ (eV) $=
512\times 10^3 E_{min}$ where $E_{min}$ is given by (6).}
\begin{tabular}{ccccccc}

\hline

& & & & $\langle f_W\rangle =2$ & $\langle f_W\rangle =10$ & 
$\langle f_W\rangle =22$ \\

$L$ & $B_0$  & $\beta_A$ &$E_{min} (eV)$ & $\Delta B$ & $\Delta B$ &
$\Delta B$ \\

\hline

3   & 11.6 & $2.67\times 10^{-2}$ & 183 & 427 & 191 & 129 \\
4   & 4.85 & $1.13\times 10^{-2}$ & 33  & 222 & 99  & 67 \\
5   & 2.50 & $5.75\times 10^{-3}$ & 9   & 147 & 66  & 44  \\
6.6 & 1.10 & $2.53\times 10^{-3}$ & 2   & 87  & 39  & 26  \\
\hline
\end{tabular}
\label{table1}
\end{table}

\clearpage

\begin{table}   
\caption{
The required average wave amplitudes $\Delta B$ (nT), as calculated from
(\ref{23}), that correspond to an acceleration timescale $T_A$ of about
10 hours, at the given locations $L$, for the given values of the spectral
index $q$, and the minimum wave frequencies
$f_{min}=2,\,10$ mHz. The parameter $\beta=0.6$. }

\begin{tabular}{cccccccccc}

\hline

 & \multicolumn{4}{c}{$f_{min}=2$ mHz}& ~ &
\multicolumn{4}{c}{$f_{min}=10$ mHz}\\

\cline{2-5} \cline{7-10}   

$L$ & $q=1.5$ & $q=5/3$ & $q=2.5$ & $q=4.0$ &
   & $q=1.5$ & $q=5/3$ & $q=2.5$ & $q=4.0$ \\

\hline

3   & 11  & 24 & 201 & 174 & & 7.2 & 14  & 90 & 78 \\
4   & 9.5 & 18 & 115 & 100 & & 6.3 & 11  & 51 & 45 \\
5   & 8.8 & 15 & 76  & 66  & & 5.9 & 8.8 & 34 & 30  \\
6.6 & 8.1 & 12 & 47  & 40  & & 5.4 & 7.1 & 21 & 18  \\

\hline
\end{tabular}
\label{table2}
\end{table}

\clearpage

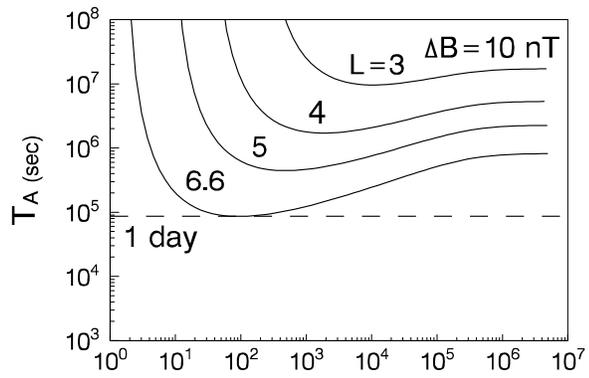
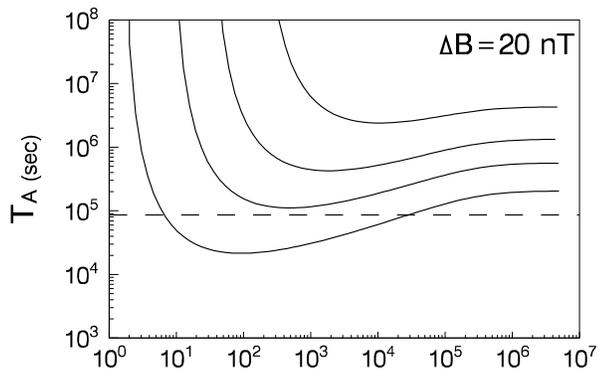
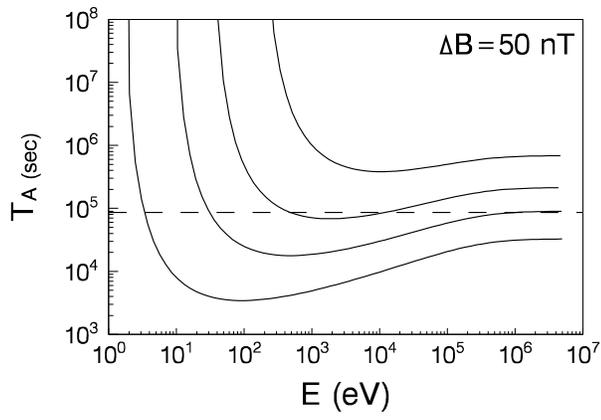
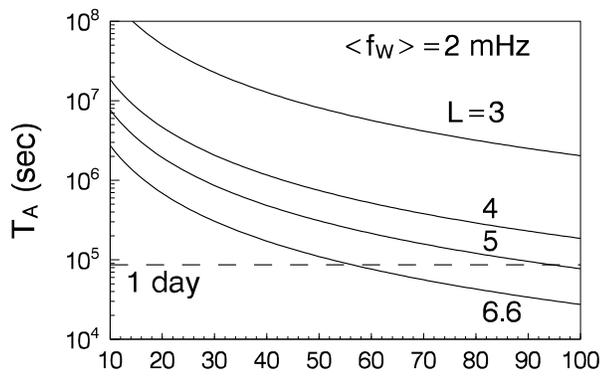
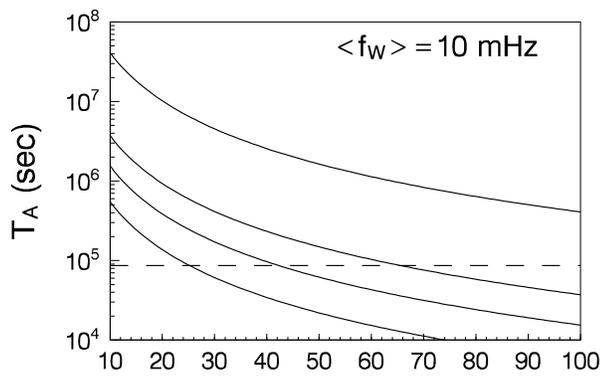
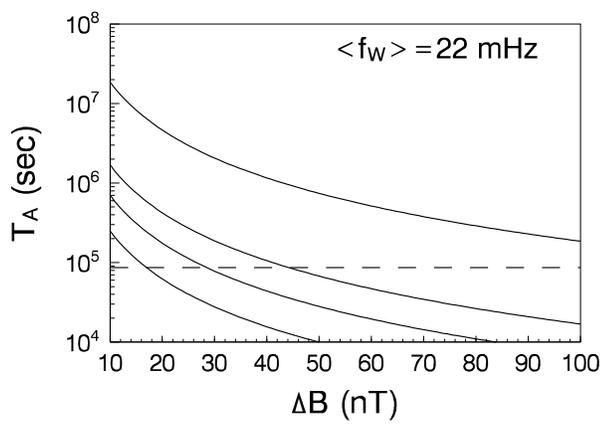
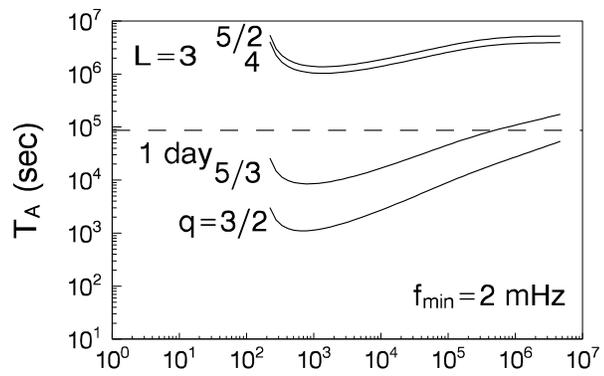
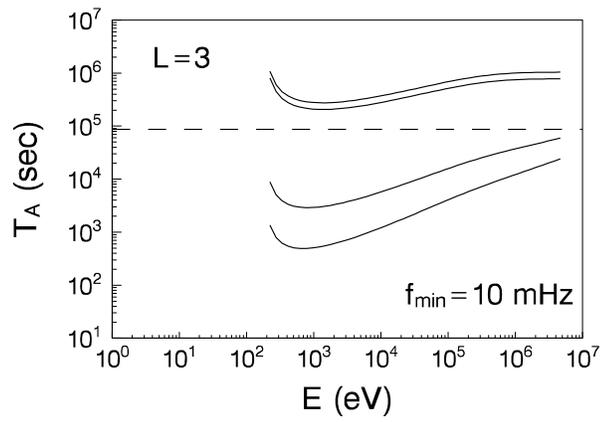
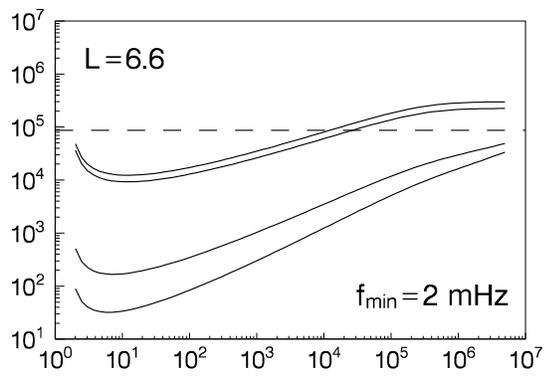
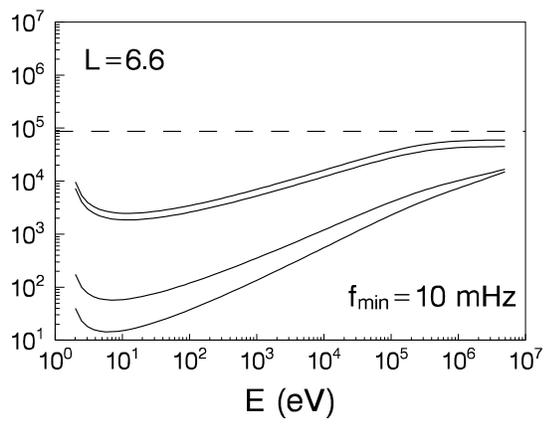
\end{document}